\documentclass[reprint,amsmath,amssymb,prl]{revtex4-1}

\usepackage{graphicx}
{}
\usepackage{hyperref}

\newcommand{\taa}{\lambda}
\newcommand{\tac}{\lambda_3}
\newcommand{\comment}[1]{}

\begin{document}

\title{
Stability of AdS in Einstein Gauss Bonnet Gravity}
\author{Nils Deppe}\email{nd357@cornell.edu}
\altaffiliation[Current address:]{~Physics Dept., Cornell University}
\affiliation{Physics Department, University of Winnipeg}
\author{Allison Kolly}\email{allisonkolly@gmail.com}
\affiliation{Physics Department, University of Winnipeg}
\author{Andrew Frey}\email{a.frey@uwinnipeg.ca}
\affiliation{Physics Department, University of Winnipeg}
\author{Gabor Kunstatter}\email{g.kunstatter@uwinnipeg.ca}
\affiliation{Physics Department, University of Winnipeg}
\date{\today}
\begin{abstract}
Recently it has been argued that in Einstein gravity Anti-de Sitter 
spacetime is unstable against the
formation of black holes for a large class of arbitrarily small
perturbations.  We examine the effects of including a
Gauss-Bonnet term.
In five
dimensions, spherically symmetric Einstein-Gauss-Bonnet gravity
has two key features: Choptuik scaling exhibits a radius gap, and the 
mass function goes
to a finite value as the horizon radius vanishes. 
These suggest that black holes will not form dynamically if the total 
mass/energy content of the spacetime is too small, thereby restoring 
the stability of AdS spacetime in this context.   
We support this claim with numerical simulations and uncover a rich 
structure in horizon radii and formation times as a function of perturbation
amplitude. 
\end{abstract}
\maketitle

\section{Introduction}
 
Anti-de Sitter (AdS) spacetime has been shown to be unstable against 
the formation of
black holes for a large class of arbitrarily small perturbations,
except for
specific initial data
\cite{Bizon2011,Jalmuzna2011,Buchel2012,Bizon2013,Buchel2013,Maliborski2013,Maliborski2013a,Maliborski2013b,Maliborski2014}. 
Given the interpretation of black hole formation as thermalization in the
AdS/conformal field theory (CFT) duality, 
the questions of stability and turbulence of AdS are very
important. 
The instability is apparently due to a subtle interplay of local non-linear
dynamics and the non-local kinematical effect of the AdS
reflecting boundary. An important question therefore concerns the
dependence of the instability and turbulent behaviour on the local
dynamics. We investigate the effects of higher curvature terms, which
translate to finite $N$ and 't Hooft coupling corrections in the dual CFT.

The most tractable higher curvature term is the Gauss-Bonnet (GB) term, 
since the equations of motion 
contain only second derivatives and are readily amenable to a Hamiltonian
analysis.  Since AdS$_5$/CFT$_4$ is a primary case of interest in the context
of the AdS/CFT correspondence, we focus on 5D; the GB term, like other
curvature-squared terms, is dual to differing $a$ and $c$ central charges in
the 4D CFT.  As a result, the GB term is commonly studied in the AdS/CFT
context.

On the gravity side, the GB term changes the local dynamics
in regions of high curvature and
radically alters the critical behaviour (Choptuik scaling)
of microscopic
black hole (BH) formation \cite{Deppe2012,Golod2012}.
One interesting feature of 5D Einstein-Gauss-Bonnet (EGB) gravity 
is that the horizon radius of a static spherically symmetric BH vanishes for
a critical value of the ADM mass, so a BH cannot form dynamically 
for ADM mass less than this critical value. Such an algebraic mass gap is also 
present in the 3D Einstein gravity 
case \cite{Banados:1992wn}; nonetheless, 5D EGB gravity
differs in that the Riemann tensor is not determined by the Ricci tensor
(as opposed to 3D) and the GB term introduces a new length scale.

 Due to the reflecting boundary 
conditions at infinity in AdS spacetime, in the sub-critical region there 
are two possible endstates: a naked singularity or 
a quasi-periodic state in which the matter continues to bounce 
back and forth. It is important to determine which of these endstates 
is realized generically.

Of potentially greater interest is whether the GB term stabilizes the
spacetime above the algebraic threshold, given
evidence
\cite{Deppe2012}  that some initial data with super-critical
ADM mass still do not form black holes in asymptotically 
flat spacetime, i.e. that there is a radius gap. 
This dynamical radius gap is expected to be a feature of EGB in at least 
all odd dimensions \cite{Deppe2012} and may also be present in other 
higher curvature theories.
We confirm the presence of a radius gap and observe that in 
asymptotically AdS spacetime
it affects black hole formation even at ADM mass far above
the critical value.

In the following we present 5D numerical simulations
consistent with the conjecture that
the stability of AdS in 5D EGB gravity is restored for
arbitrarily small perturbations.  In the AdS/CFT correspondence, this would
imply that low-energy perturbations of Yang-Mills theories on $S^3$ need not
thermalize when finite $N$ and 't Hooft coupling are taken into account.
 
\section{Action and Equations of Motion}
The action for 5D EGB gravity with cosmological constant 
minimally coupled to a massless scalar is given by
\begin{eqnarray}
I&=&
\int d^5x \sqrt{-g}\left\{-\frac{1}{2}\nabla_\mu \psi \nabla^\mu \psi
+\frac{1}{2\kappa_5{}^2}\left(\vphantom{\frac{\tac}{2}}
{12}{\taa}\right.\right.\nonumber\\ 
&&\left.\left.+ {\mathcal{R}} + 
  \frac{\tac}{2}   \left[\mathcal{R}^2 - 4 \mathcal{R}_{\mu\nu}\mathcal{R}^{\mu\nu} + 
\mathcal{R}_{\mu\nu\rho\sigma}\mathcal{R}^{\mu\nu\rho\sigma}\right]\right)\right\}\!.
 \label{eq:action1}
\end{eqnarray}
We will later rescale $\psi$ to remove the Planck scale and numerical 
factors from the equations of motion.
As $R\to\infty$, any static spherically-symmetric
solution asymptotes to AdS with effective cosmological
constant $\lambda_{eff}=(1-\sqrt{1-4\taa\tac})/2\tac$.
It proves convenient to use coordinates in which the AdS scale 
$\lambda_{eff}=1$.

A Hamiltonian analysis of EGB (and more general Lovelock) gravity in the
spherically symmetric context has been carried out in
\cite{Louko1997,Taves2012,Kunstatter2012,Kunstatter2013};
due to the Hamiltonian constraint, the generalized Misner-Sharp mass function\cite{Maeda2008}
\begin{equation}
\mathcal{M}= \frac{R^{4}}{2}\left[\taa + \frac{\left(1-R_{,\mu}R^{,\mu}\right)}{R^2}
+\frac{\tac}{R^4}\left(1-R_{,\mu}R^{,\mu}\right)^2\right],\label{eq:M}
\end{equation}
gives the energy due to matter within radius $R$ and 
asymptotes to the ADM mass at $R\to\infty$ \cite{Nozawa2008}.
In terms of the mass function, the horizon condition
$(R_{,\mu}R^{,\mu})|_{R_H}=0$ is
\begin{equation}
\mathcal{M}(R_H)=\frac{1}{2}\left[\taa R_H^{4} + R_H^{2}+{\tac}\right]\ ,
\end{equation}
which implies that $R_H\to 0$ as $\mathcal{M}(R_H)\to M_{crit}\equiv \tac/2$
even in the dynamical context. This
suggests that it is impossible to form a BH when the ADM mass is 
less than this critical value. This feature is specific to 5D EGB,
as it depends critically on the exponent of $R_H$ in the third
term of the mass function.

To connect more readily to previous literature, we work in Schwarzschild-like
coordinates with metric
\begin{equation}
ds^2 =  R_{,x}\left(-Ae^{-2\delta} dt^2 + A^{-1}dx^2 + \frac{R^2}{R_{,x}} 
d\Omega_3\right)
\label{Eq:bizonMetric}
\end{equation}
and spatial coordinate
$R= \tan (x)$.  In future work, we will consider AdS gravitational collapse
in flat-slice coordinates, which are useful for studying scaling and 
singularity formation since they allow evolution
past apparent horizon formation.

The resulting first order equations of motion are
\begin{eqnarray}
\Phi_{,t}&=& \left(Ae^{-\delta}\Pi\right)_{,x}
\label{eq:eom1}\\
\Pi_{,t}&=& \frac{3}{\sin(x)\cos(x)}Ae^{-\delta} \Phi + 
\left(Ae^{-\delta}\Phi\right)_{,x}
\label{eq:eom2}\\
\delta_{,x}&=& - \frac{\cos(x)\sin^3(x) (\Pi^2+\Phi^2)}{\left[\sin^2(x) - 
2 \lambda_3\left(A-\cos^2(x)\right)\right]}
\label{eq:eom_delta}\\
\mathcal{M}_{,x} &=& \frac{A}{2}\tan^{3}(x) (\Pi^2+\Phi^2)
\label{eq:ham1}\\
A &=&1+ \frac{\sin^2(x) (1-2\lambda_3)}{2\lambda_3}\nonumber\\
&&\times\left[1 -  \sqrt{1 + \frac{8\mathcal{M}\lambda_3}{(1-2\lambda_3)^2\tan^{4}(x)}}
\right]\ .
\label{eq:ham2}
\end{eqnarray}
Here, $\Phi=\psi_{,x}$ and $\Pi$ is conjugate to $\psi$.
In this parameterization the horizon condition is $A=0$.

The boundary conditions at the origin are identical to those in asymptotically
flat spacetime and are well-known.  At infinity,
the boundary conditions are
\begin{eqnarray}
\Phi &=& \rho^3\left(\Phi_0 + \Phi_2\rho^2+\cdots\right), \
\Pi  = \rho^4\left(\Pi_0
+ \cdots\right),
\end{eqnarray}
where $\rho = \pi/2-x$.

We solve the system (\ref{eq:eom1}-\ref{eq:ham2})
using the method of lines. \footnote{See
Supplemental Material at
\url{http://ion.uwinnipeg.ca/~gkunstat/AdSGB2014SM/SuppMat1410.1869.pdf} 
for details on the numerical scheme and convergence data.}. 
We have verified that our code is consistently convergent, and
that conserved quantities, such as the ADM mass, remain fully
fifth order accurate throughout simulations. Additionally,
we verify that altering parts of the algorithm to higher and
lower order methods provides the expected convergence changes 

\section{Results}
\begin{figure}[!h]
  \centering
  \includegraphics[width=0.50\textwidth]{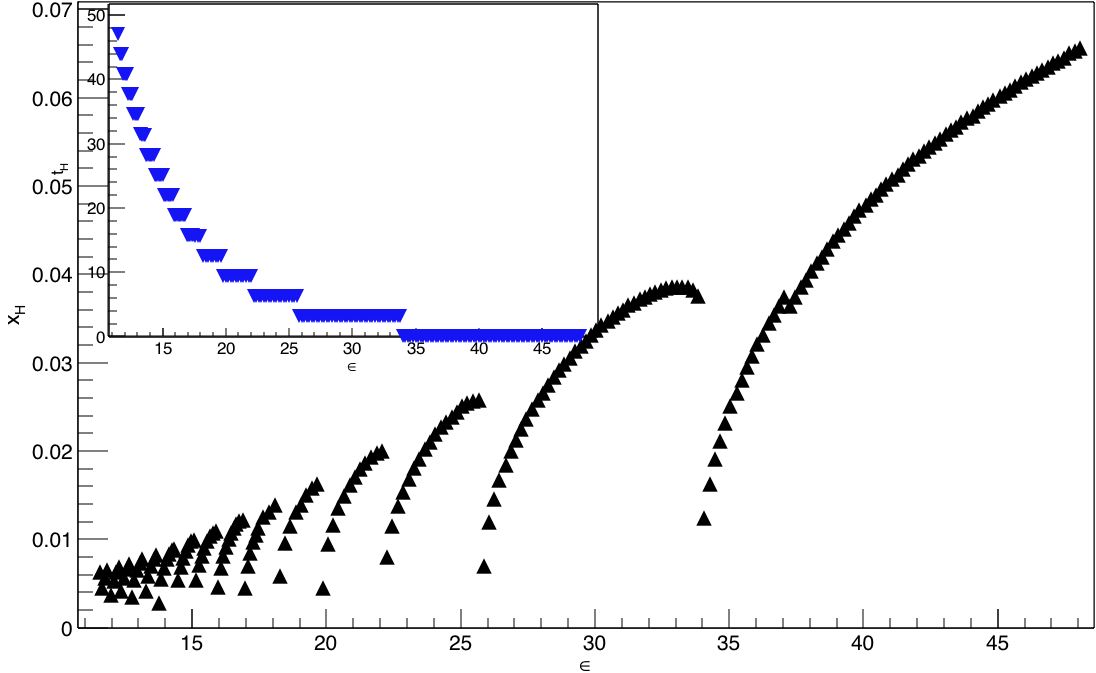}
  \caption{BH horizon radius on formation vs initial amplitude in 
Einstein gravity.  Inset: horizon formation time vs amplitude.}
  \label{fig:GR radius}
\end{figure}

In all simulations we use Gaussian initial data 
\begin{equation}
\Phi=0, \Pi = \frac{2}{\pi}\epsilon\exp\left(-\left(
\frac{2}{\pi}\frac{\tan(x)}{\sigma}\right)^2\right),
\sigma=\frac{1}{16}.
\end{equation}
Fig.~\ref{fig:GR radius} shows the
horizon radius vs amplitude for 5D 
Einstein gravity, indicating that our code gives the expected results for 
long times in this case.  Specifically, we see BH formation after the initial
pulse bounces off the AdS boundary at infinity, possibly a large number of
times. Since the coordinates break down at the horizon, the code signals horizon formation when $A(x,t)$ falls below $2^{7-k}$, where $k$ is the exponent in the number of grid points used in the simulation, i.e.~$2^{15}+1$.

The inset in Fig.~\ref{fig:GR radius} presents a plot of 
black hole formation time vs amplitude for 5D Einstein gravity. 
It illustrates that BH formation occurs soon after an integer number
of reflections from the AdS boundary (a round-trip time from origin to
boundary takes time $\pi$).
The formation time is approximately piecewise constant,
which increases exponentially in each piece as the amplitude decreases.

\begin{figure}[!h]
  \centering
  \includegraphics[width=0.5\textwidth]{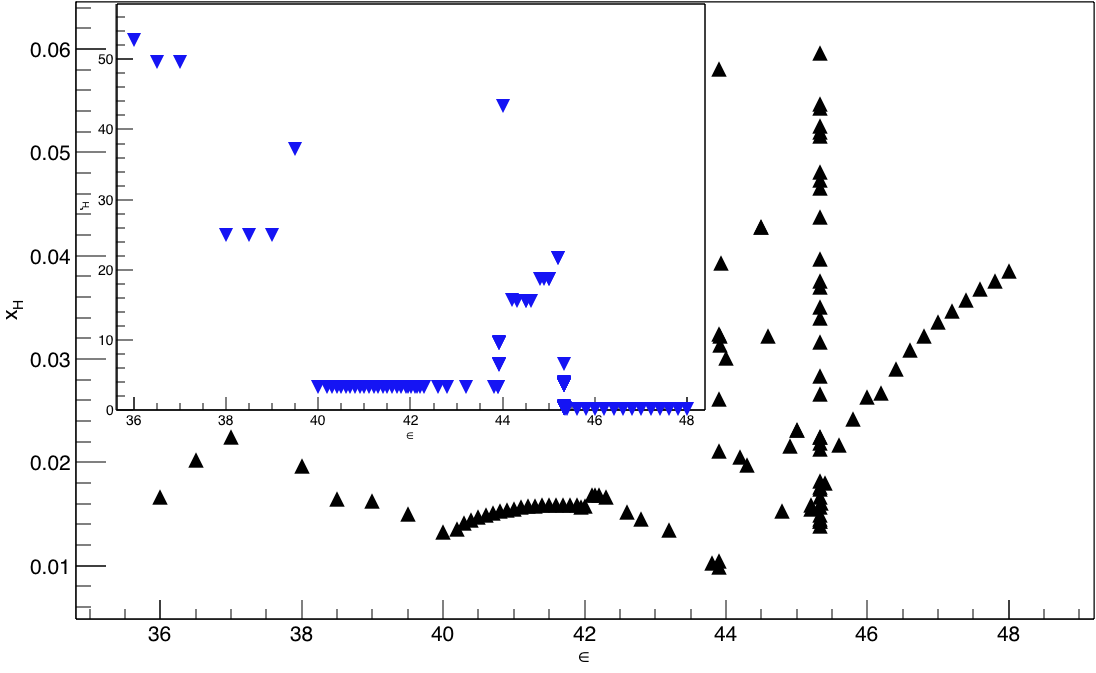}
  \caption{BH horizon radius on formation vs initial amplitude in EGB gravity,
$\tac=0.002$. Inset: horizon formation time vs amplitude.}
  \label{fig:GB radius}
\end{figure}

Fig.~\ref{fig:GB radius} shows the effect of introducing a non-zero GB
parameter, $\lambda_3 = 0.002$, for the same initial data as above. The
figures only cover the range $\epsilon=36$ to $48$ because BH formation
for lower amplitudes required many reflections and requires more computation time. The lowest
amplitude for which we successfully formed a black hole was $\epsilon=36$,
which required 24 bounces.  

The inset of Fig.~\ref{fig:GB radius} illustrates the horizon formation time vs amplitude for
the same data.  It shows that BH's form directly for large amplitudes and
transition to forming after one reflection off the boundary for 
amplitudes $\epsilon\sim 42-44$.  However, there is rich structure between 
$\epsilon\sim 44$ and 45.3, where the horizon radius and 
formation time vary unpredictably.

\begin{figure}[!h]
  \centering
  \includegraphics[width=0.5\textwidth]{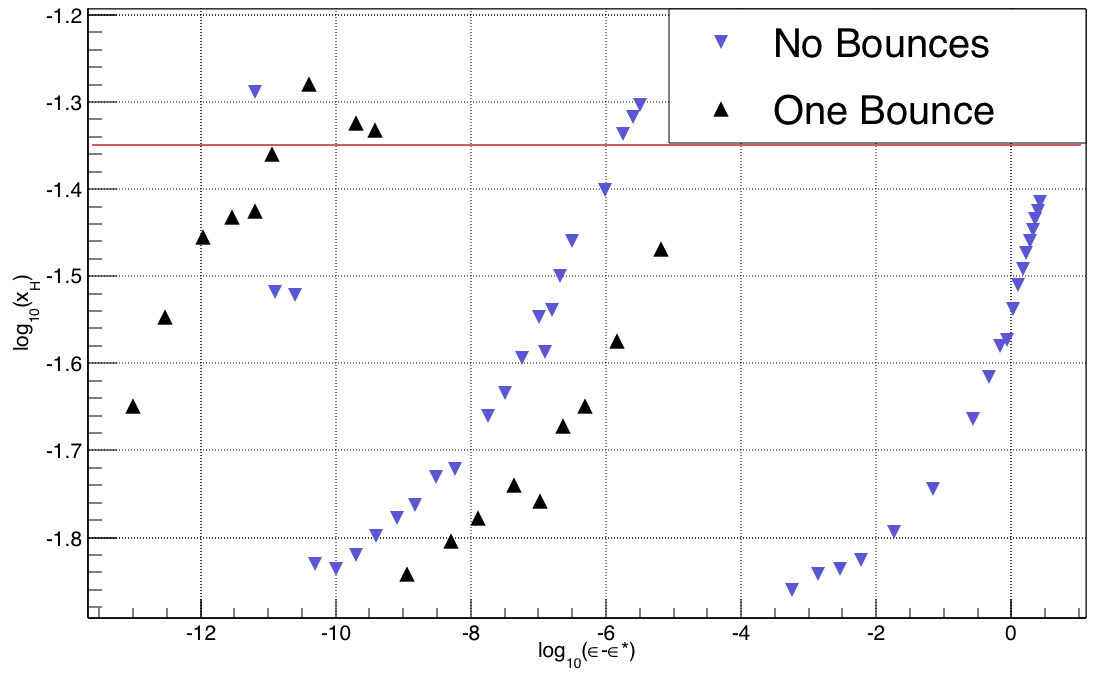}
  \caption{Scaling of horizon radius at formation after zero and one bounce for $\tac=0.002$. Both critical amplitudes are very near $45.33$.
  }
  \label{fig:gbScaling}
\end{figure}

Fig.~\ref{fig:gbScaling} shows the scaling plot as the critical amplitude $\epsilon=\epsilon^*$
for BH formation is approached after zero and one bounce. 
Whereas in Einstein gravity these would be straight lines\cite{Choptuik:1992jv} of slope
$\gamma=0.4131\pm 0.0001$ \cite{Bland2007} 
corresponding to Choptuik scaling, 
the graphs level off near $x_H \sim 0.014$ in both cases, 
suggesting the existence of a radius gap in agreement with 
\cite{Deppe2012}.

\begin{figure}[!h]
  \centering
  \includegraphics[width=0.5\textwidth]{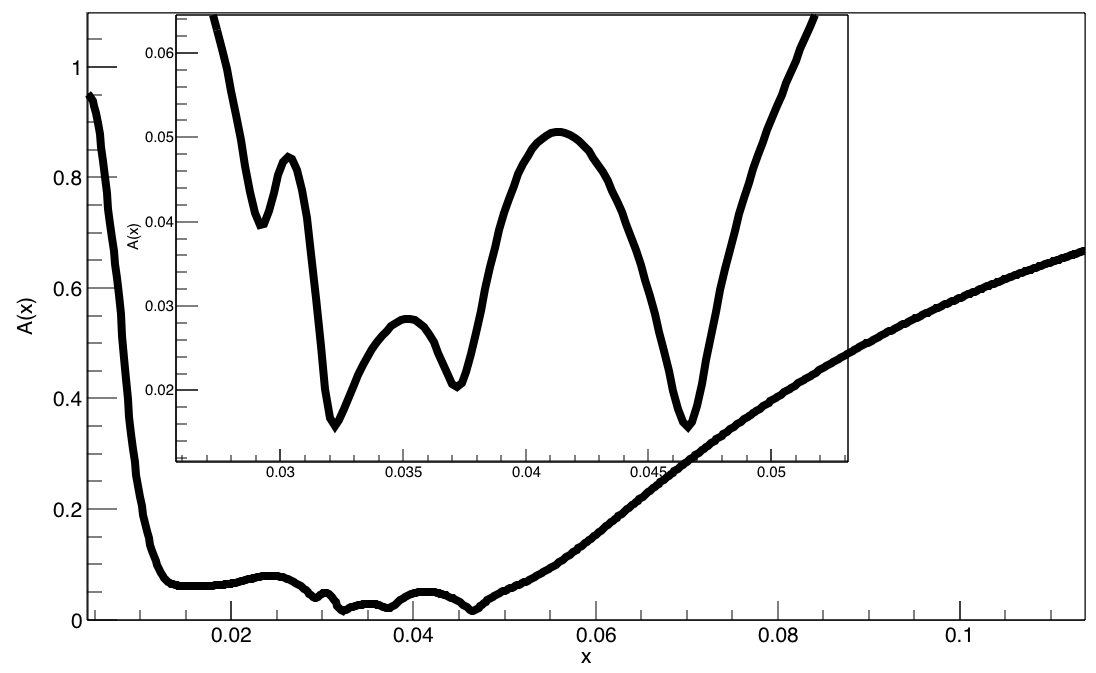}
  \caption{Metric function $A$ just prior to horizon formation for
$\epsilon=45.33143351875$.  Inset: zoomed to show local minima.}
  \label{fig:GB AH_full}
\end{figure}

Another feature of both sets of data is a jump in horizon radius as
the amplitude is lowered. This can be understood by considering the
horizon function, $A(x,t)$. In particular, when the horizon radius
gets small, $A(x,t)$ flattens out near horizon formation  and
additional minima (see Fig.~\ref{fig:GB AH_full}) appear. The jump in
horizon radius occurs as an outer minimum ``overtakes'' the inner ones
in reaching the value that signals horizon formation in the code
first. This indicates that the scalar pulse forms
multiple thick shells interior to the outer minimum.

To address the question of the endstate for ADM mass below
$M_{crit}$, we simulated an amplitude $\epsilon=20$,
where $\epsilon_{crit}=21.86$ corresponds to $M_{crit}$.  Without the GB term
this amplitude results in black hole formation after three bounces.
In the present case the simulation was continued to $t=200$, corresponding to over 60
bounces, 
 with no horizon formation. The dynamics of the pulse as it bounces back and forth is
quite intricate \footnote{Movies of sub-critical collapse simulations
are available at 
\url{http://ion.uwinnipeg.ca/~gkunstat/AdSGB2014SM/}.}.

\begin{figure}[!h]
  \centering
  \includegraphics[width=0.5\textwidth]{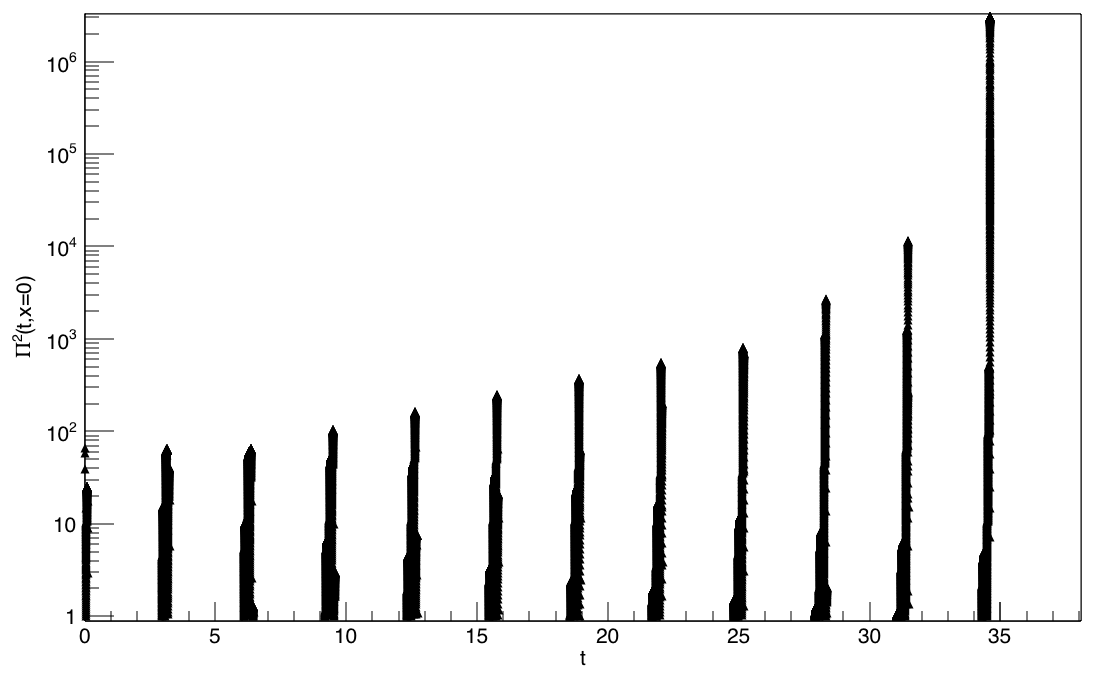}
  \caption{$\Pi^2(x=0,t)$ in Einstein gravity for $\epsilon=12.7$}
  \label{fig:GR Long}
\end{figure}

\begin{figure}[!h]
  \centering
  \includegraphics[width=0.5\textwidth]{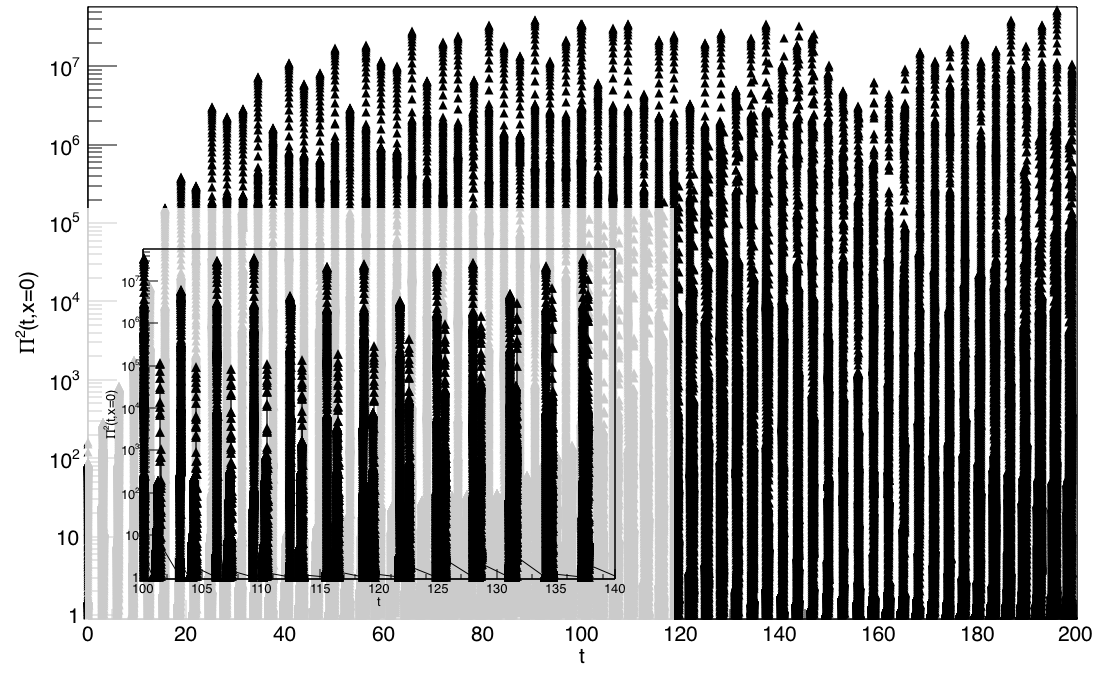}
  \caption{$\Pi^2(x=0,t)$ in EGB gravity for $\epsilon=20$. Inset: zoomed
to show peaks with different relative phases.}
  \label{fig:GbPiInset}
\end{figure}

Comparison to Einstein gravity is instructive.  Fig.~\ref{fig:GR Long} 
graphs $\Pi^2$ at the origin, which is proportional to the trace of the
stress tensor, for $\epsilon=12.7$ in Einstein gravity.
The tendency of the scalar pulse to get more
 concentrated, or focused, at the origin after each
bounce from the boundary is apparent in the steadily increasing peak value
of $\Pi^2$.  Fig.~\ref{fig:GbPiInset} graphs $\Pi^2$ for $\epsilon=20$ 
in EGB gravity. In contrast, the pattern is irregular, and
there is no apparent tendency to focus.

From the inset in Fig.~\ref{fig:GbPiInset}
one can see that there are multiple peaks of $\Pi^2(x=0)$.
This agrees with our observations from animations that the GB term
causes the original pulse to break up into multiple smaller pulses,
which then propagate through the spacetime. The GB term causes delays
in the implosions resulting in a slightly different phase for the
different pulses. We have observed that BH's form when a sufficient
number of these pulses are within the horizon radius at the same time.
Interestingly, this does not necessarily translate into the curvature
being large at the origin.

Additionally, the energy spectrum of the $\epsilon=20$ pulse shows no
evidence of a turbulent cascade of energy to higher frequencies as time
passes \cite{Note2}.  
This provides some support to the notion that the system settles
into a smooth quasi-periodic state, however more simulations are
necessary to draw a definitive conclusion.

The above results are in stark contrast with what is seen
in the 3D case where an algebraic mass gap is also present.
In 3D Einstein gravity, there is no lower bound on the BH radius
\cite{Pretorius2000}, whereas the BH radius is bounded below in the present
case.  This behavior seems closely related to the complex structure seen
in Fig.~\ref{fig:GB radius}. 
Further, the energy spectra for sub-critical collapse in 3D does not share this characteristic behaviour
\cite{Bizon:2013xha,Jalmuzna:2013rwa}.
\comment{ and 5D EGB do
not share characteristic behaviors.}

\comment{
In the present case we observe a type I phase transition: the black hole radius is bounded below as the critical amplitude is approached. This in turn leads to a dynamical mass gap above the algebraic one. In the 3D
GR case the transition is type II: there is no lower bound on the black hole radius\cite{Pretorius2000}.
The type I behaviour is likely closely
related to the chaotic structure observed in
Fig.~\ref{fig:GB radius}.   }
   
\section{Conclusions}

We have presented the results of numerical simulations of spherically
symmetric massless scalar field collapse in 5D AdS EGB gravity. Our data are
consistent with the conjecture that stability against small perturbations
is restored. Some speculations
are perhaps in order: After each bounce from the boundary, the 
{Einstein} term focuses the pulse of matter as it implodes at 
the origin. On the other hand, the observed dynamical radius gap leads to a 
defocusing effect that resists BH formation at small
horizon radii and allows the matter to travel to the boundary 
multiple times before BH formation.  The defocusing
effect is evident in the out-of-phase peaks in $\Pi(x=0)^2$ seen in 
Fig.~\ref{fig:GbPiInset} as well as the flattened form of the horizon 
function (Fig.~\ref{fig:GB AH_full}) in EGB gravity. This defocusing in turn 
affects the time it takes for the pulse to disperse from the origin. 
\comment{After the matter disperses from the origin the subsequent bounce(s) 
or bounces from the AdS boundary at infinity lead to a subtle interplay 
between the different timescales and dynamics at different scales. Another 
important point is that the hallmark of Choptuik-type critical collapse 
is  extreme sensitivity of the outcome (BH formation vs dispersion) to 
initial conditions. We conjecture that it is the de-focusing with its 
comcomittant interplay between timescales  which, inlight of the inherent 
sensitivity to initial conditions, produces the complex
structure seen in Fig.~\ref{fig:GB radius}.
 Further, as the GB and {Einstein} terms compete, it is 
possible to produce very intricate wave forms.} 
Furthermore, extreme sensitivity of the outcome (BH formation vs dispersion)
to initial conditions is a hallmark of critical collapse.  This sensitivity
along with altered dispersal timescales leads to the complex structure
seen in Fig.~\ref{fig:GB radius}.
One can
speculate further that the map from amplitude to horizon formation time
may evince a fractal structure due to the interplay between Einstein and GB
dynamics at the origin. In any case, the data clearly suggest that the
GB corrections to short distance dynamics
\comment{the dynamics at short distances} inhibit the
formation of black holes and that stability may indeed be restored. Of
course, it is much more difficult to prove stability, if indeed that
is the case, than instability. We plan a detailed study of these issues in 
future work.

There are in principle an infinite number of possible higher curvature 
deformations to Einstein gravity. 
It is important to ask whether the qualitative features we observe persist 
in the more general class of deformations. In brief, 
the suppression of black hole formation in EGB 
is a consequence of the dynamical radius gap, which is indicative of a 
non-zero mass critical solution.
\comment{a type 
I Choptuik-type transition between black hole formation and dispersion at 
the origin. Such type I transitions typically occur} 
These are well-known to occur when a new length scale 
becomes relevant to the dynamics, as invariably happens 
\comment{when considering stability under} 
in gravitational collapse \comment{in theories} with higher curvature 
deformations. Thus, we expect the BH suppression to be generic in 
such theories.  Moreover, the sensitivity to initial data of critical 
collapse in combination with a radius gap should generically lead to
complex structure in pulse waveforms and BH formation time in 
higher-curvature gravities.

\comment{
Moreover, since sensitivity to initial data is a key feature of Choptuik-type 
transitions, and the de-focusing is expected to occur whenever the transition 
is type I, one expects the black hole formation time and waveforms to 
generically display the complex structure that we observe whenever there 
is a radius gap.}

In conclusion, our analysis shows that BH formation instabilities in AdS
are highly sensitive to small scale dynamics of gravity. Moreover, 
our results imply finite $N$ and coupling effects modify thermalization in
a dual field theory through the AdS/CFT correspondence.

\begin{acknowledgments}
We would like to thank A.~Buchel, M.~Choptuik, D.~Garfinkle, L.~Lehner,
A.~Rostworowski, and T.~Taves for useful conversations.  This work was 
funded in part by the Natural Sciences and Engineering Research Council 
of Canada.  Support was also provided by WestGrid (www.westgrid.ca), 
Compute Canada/Calcul Canada (www.computecanada.ca), and the Perimeter 
Institute for Theoretical Physics (funded by Industry Canada and the Province
of Ontario Ministry of Research and Innovation).
\end{acknowledgments}

\bibliography{AdS2}
\end{document}